\documentclass{article}
\usepackage{graphicx} 
\usepackage{amsmath,amsthm,amssymb}
\usepackage{siunitx}
\usepackage{wasysym}
\usepackage{etoolbox}
\usepackage{xcolor}
\usepackage{gensymb}
\usepackage{latexsym,epsfig,dcolumn,comment}
\usepackage{float}
\usepackage{caption}
\usepackage[backend=biber, style=numeric]{biblatex}
\title{Simulating Subterranean Fluid Injection through Iteration on the VirtualQuake Model}
\author{Spence Norwood, John Rundle }
\date{October 2025}

\begin{document}

\maketitle

\section{Introduction}

There is a growing need for modeling of subterranean fluid injection. Hydraulic fracturing, or fracking, has become a major source of hydrocarbons in the global economy, rapidly rising in prominence and scope. As of 2023, approximately $60\%$ of US domestic oil supply was produced in shale fields extracted by fracking\cite{MCMAHON2024925}. Yet, the technology has been linked to rise in earthquakes, even in traditionally inactive seismic regions\cite{Villa2020}. As the damage caused by these earthquakes rises, so does the requirement for better simulated data showing the effect of these extractive practices on nearby faults, to better predict the risks of future endeavors.

This need isn't limited to fracking though, as a wide range of green technologies also require fluid injection, such as carbon sequestration\cite{ZHANG2014319}, geothermal power generation\cite{GanQuanElsworthDerek}, and underground hydrogen storage\cite{hydrogenseis}. Regardless of the path the world's energy production takes, the risks of induced seismicity from these practices needs to be characterized.

\section{Iterating on VirtualQuake}

The VirtualQuake model was selected as basis for this project, serving as a template for a robust, physics based simulation of faults and their associated seismic activity. Despite that, the model has several weaknesses that require improvement. To address the shortcomings of the model and provide a more adaptable earthquake modeling software, as well as to provide needed backend for fluid modeling, the system was rebuilt using the point source solutions for an elastic half space\cite{Okada1992} , instead of the original rectangular.

The rectangular solutions are singular at the edges of the fault, which lead to instability in the old VirtualQuake system. If a fault edge connected to a fault on its surface, this would lead to an undefined value in contact with a non-singular one, which could lead to model collapse. As a result, the system was restricted to contiguous geometries, and inter-fault interactions were barred. This limited the overall usability.

In contrast, this approach tiles the fault space by assigning rectangular zones to point sources. Each point source has a vector that defines its associated fault plane and an area. As the point sources are singular at the points themselves, this allows for simulated faults to intersect and combine in any combination, as long as the sources are not directly overlapped. The approach can be generalized to non-rectangular areas, as well, so long as an associated plane still exists.

The point source solutions were implemented in Python using an object-oriented approach. Every source in the system is an instantiated implementation of an abstract class, a "Node". The abstract Node class provides specifications for stress source implementation, a uniform set of requirements and expectations for greater usages. These definitions cover common use cases, such as calculating stress or displacement on a given location or other Node. This abstract definition is then implemented in a separate file to fulfill these requirements. This approach yields benefits in several ways, one of the most useful being modularity. \cite{ObjectOrientedAdvantage}

Structures that use Nodes make no assumptions about the underlying implementation, meaning implementations are fully interchangeable in the control structures of utilizing models. A fault could be composed of tiles where every single one has different implementations and effects, and the greater model will still function. This is a degree of flexibility beyond simply changing parameters, as they can vary in terms of functional design and output as well. \cite{Fenves1990} 

Nodes being self-contained implementations of their own effects and that of their effects on others allows for them to be easily taken from the greater algorithm and applied to other functions and needs. Current implementations only require basic python packages, without reliance on the greater control structures described in this package. The structures can be easily reused in other formats and applications.

The sources are initialized with parameters that define the local conditions, as well as a full set of f which define and calculate the source's behavior.

When using point sources to approximate a rectangular solution, eventual convergence is guaranteed with sufficient density of increasingly smaller tiles, as this mimics the mathematical convergence of the integration that derived the rectangular equations. However, there is a need to characterize the required density for a reasonable approximation. The less dense the tiling, the faster the computation.

\begin{figure}[H]
    \centering
    \includegraphics[width=9cm]{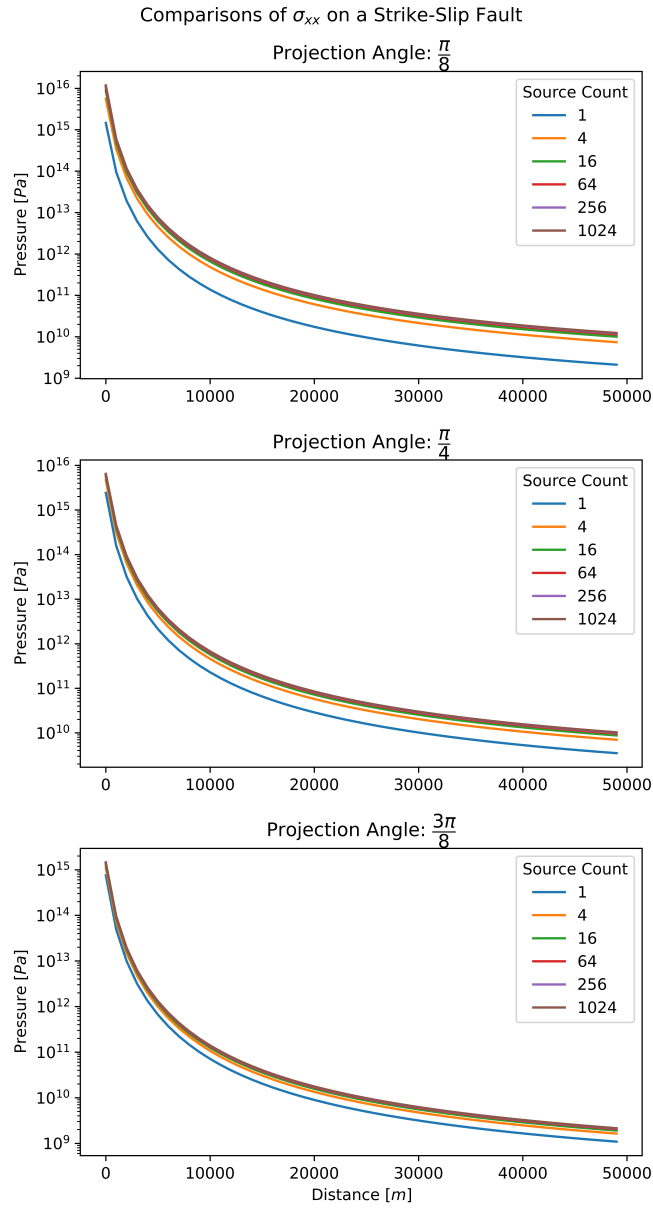}
    \caption{The convergence of a tiled point source system with increasing source density} 
\end{figure}

In figure one, a 1km square strike-slip fault segment is tiled with an increasing density of point sources. Measurements of the $xx$ component of the stress tensor are taken at increasing distances away from the fault, at several sample angles. The convergence to common solutions is rapid.

\begin{figure}[H]
    \centering
    \includegraphics[width=9cm]{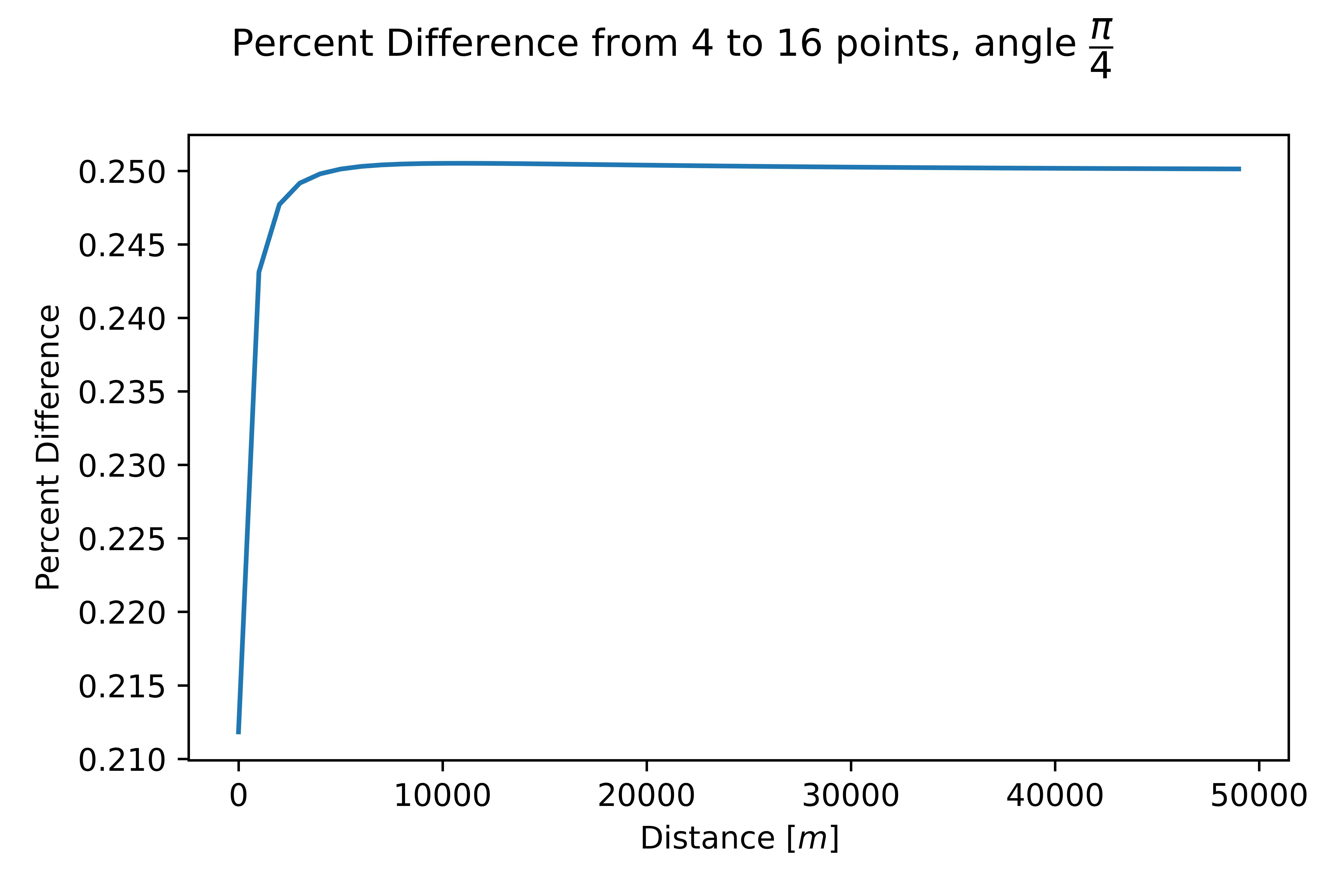}
    \caption{Percent difference in magnitude between a tiled fault with 4 and 16 sources, respectively.} 
\end{figure}

The most rapid change occurs when going from one to four points, and after that, change is limited. Increasing the density by another factor of four, from 4 to 16, results in less than half a percent of difference on average. This allows for relatively sparse tiling regimes for size regions of interest, with near equivalent output, a boon for performance.

The usage of point sources over rectangular tiles has another benefit, in that the points can be distributed arbitrarily, in any orientation. Modeling faults as large contiguous rectangular segments is serviceable for approximate models, but fault surfaces have been shown to exhibit self similar geometry, a scale invariant repetition that signifies fractal behavior. This variation has important effects on the local shear friction and permeability\cite{PowerWillamTullisTerry}, and being able to replicate that is important. The new model has support for generation of fault plane segments, tiled with density selected by user specification, but also, taking advantage of the greater freedom of distribution of point sources, has the option to include fractional Brownian noise in the geometry generation.

\begin{figure}[H]
    \centering
    \includegraphics[width=12cm]{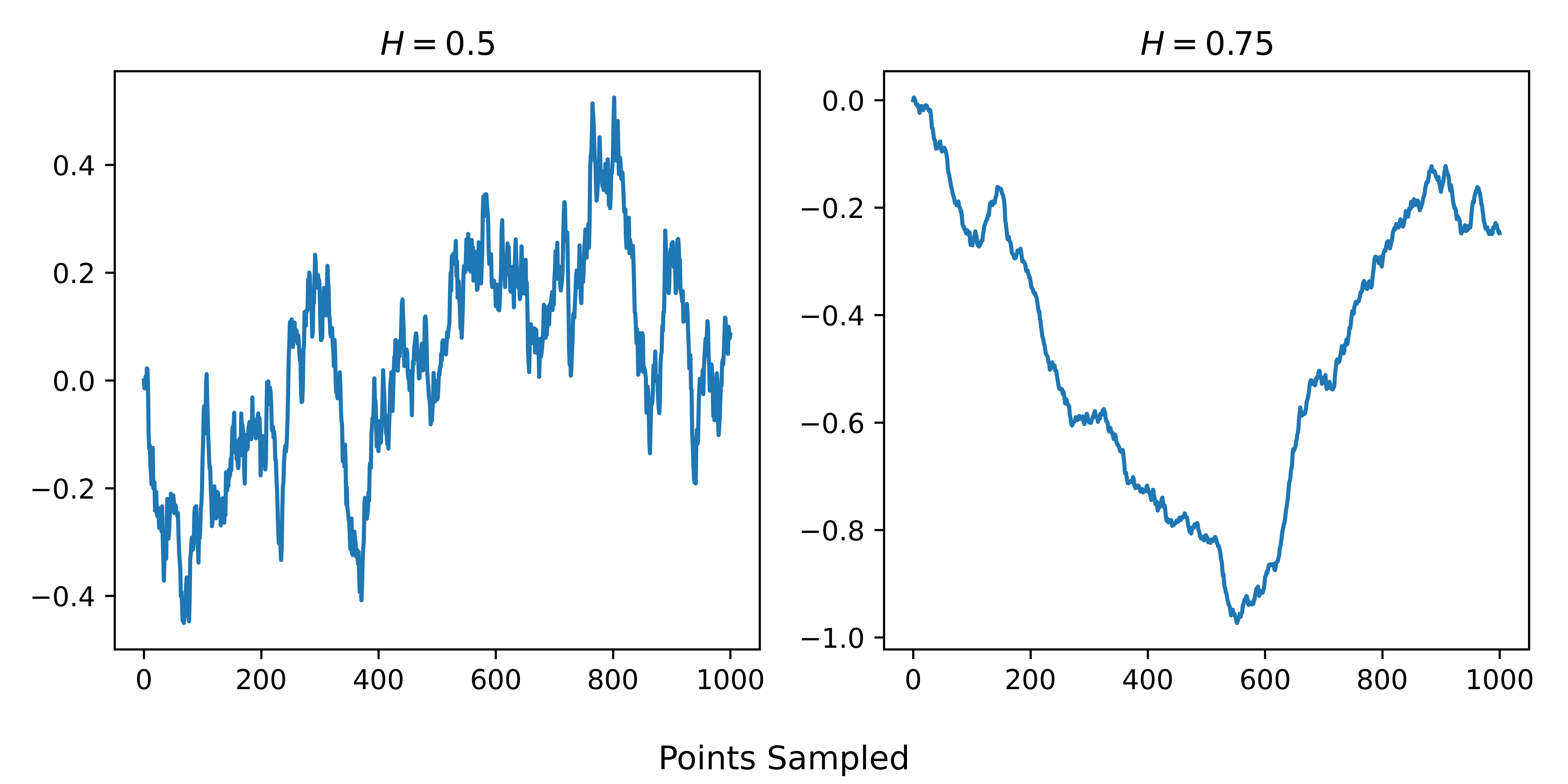}
    \caption{The behavioral differences of Fractional Brownian Motion with varying Hurst Parameter} 
\end{figure}

Brownian motion is a model designed to describe the motion of a particle suspended in a medium, and is a fully random model from iteration to iteration. Fractional Brownian motion is the generalized form, differing by allowing for the existence of a correlation between the function values, where the covariance is

\begin{equation}
    \frac{1}{2}(|t|^{2H} + |s|^{2H} - |t-s|^{2H}
\end{equation}

between any two times $t$ and $s$, and $H$ is a constant called the Hurst parameter, between 0 and 1. At $H=1/2$, the process is again fully independent from interval to interval, but is correlated positively or negatively as H approaches 1 or 0. Fractional Brownian motion is self similar and is a fractal process. Earth surfaces have been shown to be modelable by this process,\cite{HUANG1988223} with H values from .75 to .8 being typical.\cite{nhess-8-657-2008}.

\begin{figure}[H]
    \centering
    \includegraphics[width=9cm]{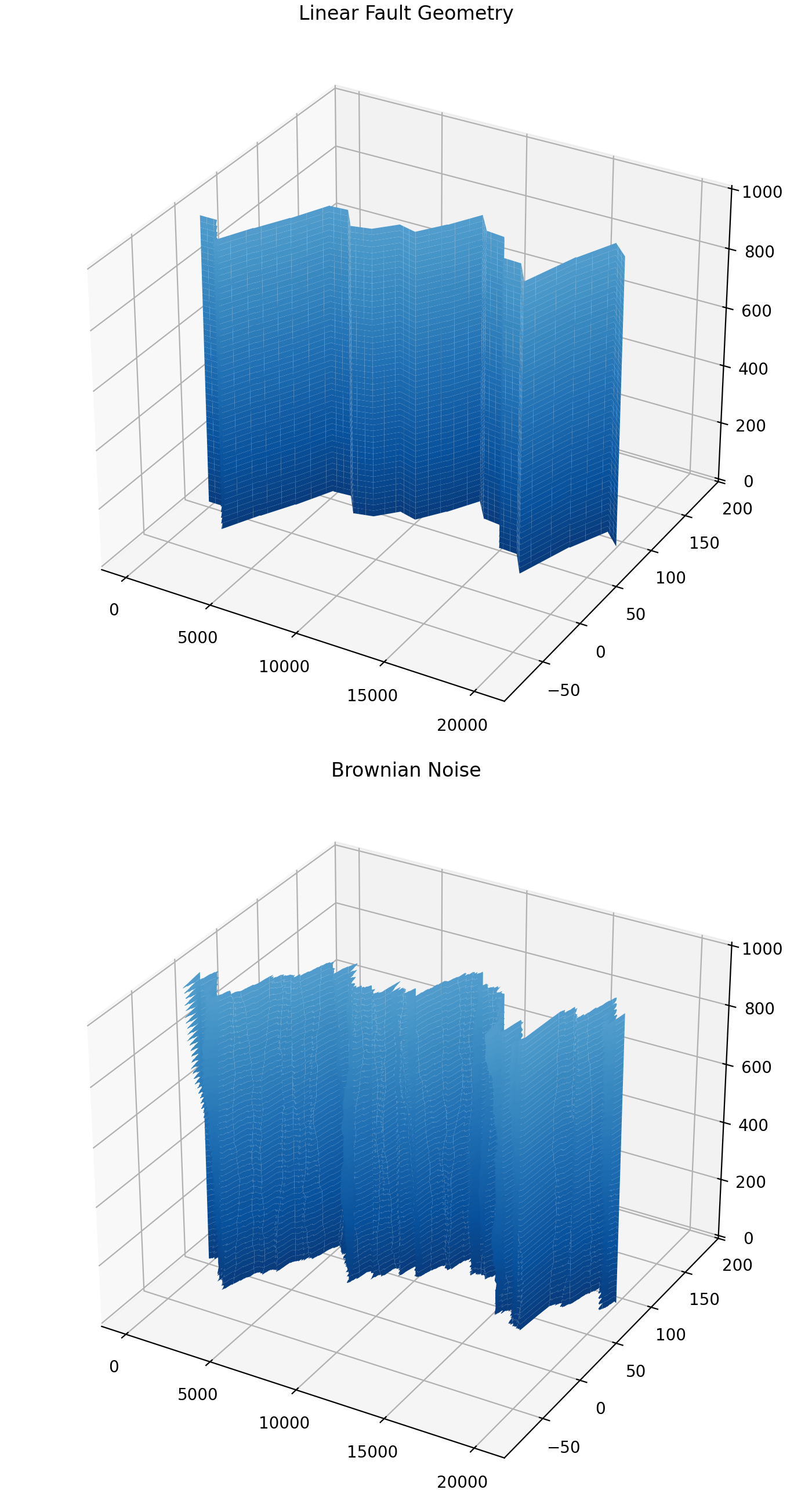}
    \caption{The result of a two-dimensional Brownian transformation on a sample fault geometry} 
\end{figure}

To apply this to the model and a given plane element, a two dimensional Fractional Brownian Field is created for a given Hurst parameter\cite{2DBrownianCode}. This distribution is sampled at the corresponding point sources making up the plane and the value multiplied by a scaling magnitude term to match the scale of the element. These values are applied as a transformation on the point grid, displacing it from the uniform generation pattern. Strike, slip, and dip terms are interpolated to fit the new geometry, and fed into the model as normal. This process allows for a fast and simple modification to deeply increase the complexity of the fault, and the variation in potential behavior.

To begin instantiation, the fault geometry must be defined in advance. The system allows for free input of Nodes in any configuration, but has default support for generating tiled plane segments. After all static fault elements are defined, and their corresponding Nodes structure created, the fault configuration is locked, disallowing later additions.

Then, for every Node in the fault system, a stress tensor is calculated for the effect of that Node on all others, as well as itself. These tensors are projected onto the planar surface of the affected Node, and decomposed into shear and normal components. The results of these calculations are stored in two $NxN$ matrices. With the same method as VirtualQuake, these Green's functions are used to calculate the resultant shear and normal stresses on a node.

\begin{equation}
    \label{eq:stressef}
    \begin{aligned}
        \tau _{A} = \sum_{B} T_{s}^{AB} s_{B}(t)\\
        \sigma_{A}=\sum_{B} T_{n}^{AB} s_{B}(t)
    \end{aligned}
\end{equation}

Summing over the contributions from all Nodes, the aggregate values of shear and normal stress on a Node A are found, where $T_{s}^{AB}$ and $T_{n}^{AB}$ are the projected effects from a Node B on the normal and shear stresses of Node A, and $S_B$ is the slip magnitude on that node. The greater the slip on a node, the larger the projected stress from the Green's functions. The time progression of the system is simple, with slip being advanced on each node linearly with time, according to a specified slip rate.

\begin{equation}
    \label{eq:cff}
    CFF^{A}(t) = \tau^{A}(t) - \mu^A(\sigma^A(t) +\rho g d)
\end{equation}

The criterion for a fault segment rupture and the beginning of an earthquake is a simple one, the Coulumb Failure Function\cite{https://doi.org/10.1029/JB074i022p05343}, $CFF$. Shear and normal forces on a potential rupture plane are opposed, the normal forces holding the segment in place from static friction, and the shear forces attempting to cause a shift. The only forces considered are the aggregate shear and normal forces from the internode effects, and the hydrostatic pressure of the local environment, $\rho g d$. As the shear forces rise over time, as fault slip values increase, the function value approaches zero, which signals an element failure.

\begin{equation}
    \Delta s = \frac{1}{K_L}(\Delta\sigma - CFF)
\end{equation}
When a node has a CFF of zero or greater, it is flagged and the slip is lowered on the node by an amount $\Delta s$, where $K_L$ is a self-stiffness term of the element itself, and $\Delta \sigma$ is a characteristic stress drop defined in terms of seismic variables of the system. The most positive the $CFF$ of an element, the greater the corresponding slip drop. 

\begin{equation}
    \label{eq:seismicconst}
    \begin{aligned} 
        \sigma_{A} = -2\mu \Delta s^{c}\pi \sqrt{2A}\frac{2-\nu}{{1-\nu}} \\
        \Delta s^{c} = 10^{\frac{3}{2}(M^{c} + 10.7} \\
        M_{c} = 4 + log_{10}(A) + s_d 
    \end{aligned}
\end{equation}

For further details on these constants and their constituent terms, reference the VirtualQuake manual.

After the slip is lowered on the initial failure element, the stress and CFF values are recalculated in the new state. This can lead to additional element failures, which are then lowered in turn, this process continuing until all elements have a $CFF$ lower than zero. This cascade of failures is an earthquake event, and the sum of all rupture elements and their relative slip change determines the seismic magnitude of the event.

The fault geometry is assumed to be static in time, unchanging as the seismic system evolves. To maintain this, instead of true slip, a backslip approach is used, where a reverse displacement is applied to the element. Then, when the element ruptures, it returns towards the equilibrium position.\cite{npg-18-955-2011}

\section{Introducing Fluid Injection}

This system emulates the capability and behavior of the VirtualQuake model, but must be extended to account for fluid injection. Injected fluid does not have a static geometry and the pressure exerted by a fluid is not well modeled with rectangular fault equations. To represent a region of pressurized fluid, point sources representing inflationary elastic half space solutions will be distributed according to an invasion percolation algorithm. 

Invasion percolation is a variant of percolation originally designed to model the displacement of a fluid displacing another in a porous medium.\cite{Wilkinson_1983} The process begins with the invading fluid occupying one cell in a gridded lattice. Each cell is connected to all adjacent cells with a randomized bond strength, representing the pores and their resistance. With each iteration, the lowest strength bond is selected and broken, and the invading fluid occupies the selected cell. 

This algorithm grows in clusters called bursts, periods of growth where only bonds are broken existing under a given threshold of bond strength. These periods of growth are often clustered in the same local vicinity, newly exposed weak bonds explored before revisiting older, stronger connections. \cite{PhysRevLett.61.2117} In the non-trapping version of this algorithm, where encircled regions of unexplored space are still open to fluid replacement, these bursts can be mapped to fit the Gutenberg-Richter relation, and have been suggested as a modeling mechanism for micro-seismicity. \cite{PhysRevE.89.022119}

As the percolation algorithm progresses, each occupied cell is mapped to a location in real space and an inflationary solution Node is initialized there. Unlike before, the stress effects of these inflationary Nodes on each other is not calculated, only the effect on the fault elements. These new Green's solutions are appended onto the fault's two interaction matrices, resulting in their contribution being included in future calculations of the Coulumb Failure Function, and thus potential seismicity.

By default this injection process is in three dimensional lattice. The bonds have bias factors that can be set to increase the likelihood of weaker or stronger bonds in any given axis. This allows for the percolation growth to be responsive to local conditions in the ground, or toward sets growth patterns. When faults are activated and under induced pressure, there are often drastic increases in local permeability along the axis of fault systems, up to several orders of magnitude. Aligning the lattice grid along the axis of the fault, and lowering the relative strength of the bonds in that direction allows the percolation to grow in a more realistic way for near fault injections.

Placing inflationary stress sources to simulated pressurized fluid requires a mapping between the seismic moment scaling term of the Okada solutions, and the variable pressure of the fluid as it is injected and flows.

\begin{equation}
    W^{ijk} = P^{ijk}d^3\phi
\end{equation}

As part of the earlier spatial mapping, a fixed distance $d$ is assigned to connections between lattice sites. Each lattice encompassing a cubical region of space and has an associated fixed volume. By multiplying that with the lattice site pressure, $P^{ijk}$, and porosity $\phi$, the work of adding the fluid is derived. This serves as an analogue to the effective seismic moment of the fluid.

Pressure varies from the injection site outward. Pressurized injection has the potentially to be highly non-laminar when the fluid velocity is high\cite{Zeng2006}, so the Forchheimer equation must be used to describe the pressure gradient $\nabla P$.

\[ \nabla p = -\frac{\mu}{k} v - \frac{\rho}{k_1} v^2\]

where $\rho$ is the fluid density, $v$ is the superficial velocity, $k$ is the permeability, $\mu$ is the fluid viscosity, and $k_1$ is the inertial permeability. To find the superficial velocity of the fluid, an estimate of the cross-sectional flow area must be made for arbitrary distances from the injection side. This calculation is not trivial, as the growth of invasion percolation exhibits fractal behavior\cite{PhysRevA.42.7496}, so surface area does not have a well-defined value. An estimation of this value is calculated by tracking the accumulated surface area of the occupied cells, and then rescaling it according a simulation constant.

\begin{figure}[H]
    \centering
    \includegraphics[width=9cm]{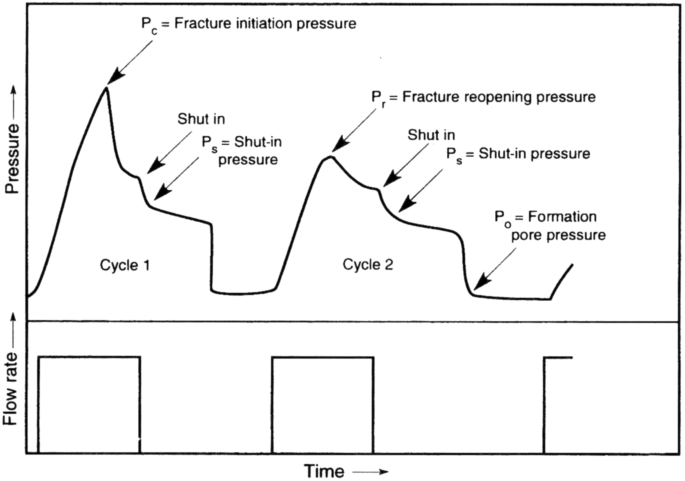}
    \caption{Pressure dynamics at the beginning of a series of fracking cycles\cite{amadei1997rock}} 
\end{figure}

This process can be tailored to one of the most common types of fluid injection, fracking. Fracking, the common name for hydraulic fracturing, injects high pressure fluid to induce a fracture through a resource rich region to allow for cheaper and faster extraction. This process most commonly involves a horizontally drilled injection tunnel, which is slowly grown as the fracture is expanded.

This process begins in a period of increasing pressurization in the injection borehole, until an initial fracture is created. This pressure is commonly referred to as the breakdown pressure. After this initial breach in the local rock formation is created, flow of the fracking fluid begins and the pressure at the injection well drops to a lower but relatively stable value, as the fracture is grown and propagated for a selected time of total injection. After this period is reached, the inject flow is halted at the marked Shut-in time\cite{WU2020103185}. 

In active work sites, this period is brief, as the fracture is prepared for further drilling and expansion, to grow the fracture along a horizontal axis to the earth. The fracture is repressurized, albeit to a lower magnitude than the initial breakdown pressure, and the fault is grown further. The pressure required to grow the faults in later cycles is lower than the initial breakdown pressure due to weakening of the formation from the mechanical stress and damage.

Eventually fracturing on an area is halted, and the borehole is sealed for a longer, potentially indefinite duration. The internal pressures of the fluid will stabilize, local gradients evening as the injection disturbance dies away. Over time, the pressurized fluid inside the fraction will leach into the surrounding fluid reservoir and pressures will equalize.

This process though, can potentially be very slow. Shale, a common strata encountered in fracking, is extremely impermeable in its unfractured state, with permabilities in the range of $10^{-18}$ to $10^{-21} m^{2}$\cite{ShaleFacts}. This leads to pockets of elevated pressure existing for timescales far in excess of the rate of fracking operations.

\begin{figure}[H]
\centering
    \includegraphics[width=14cm]{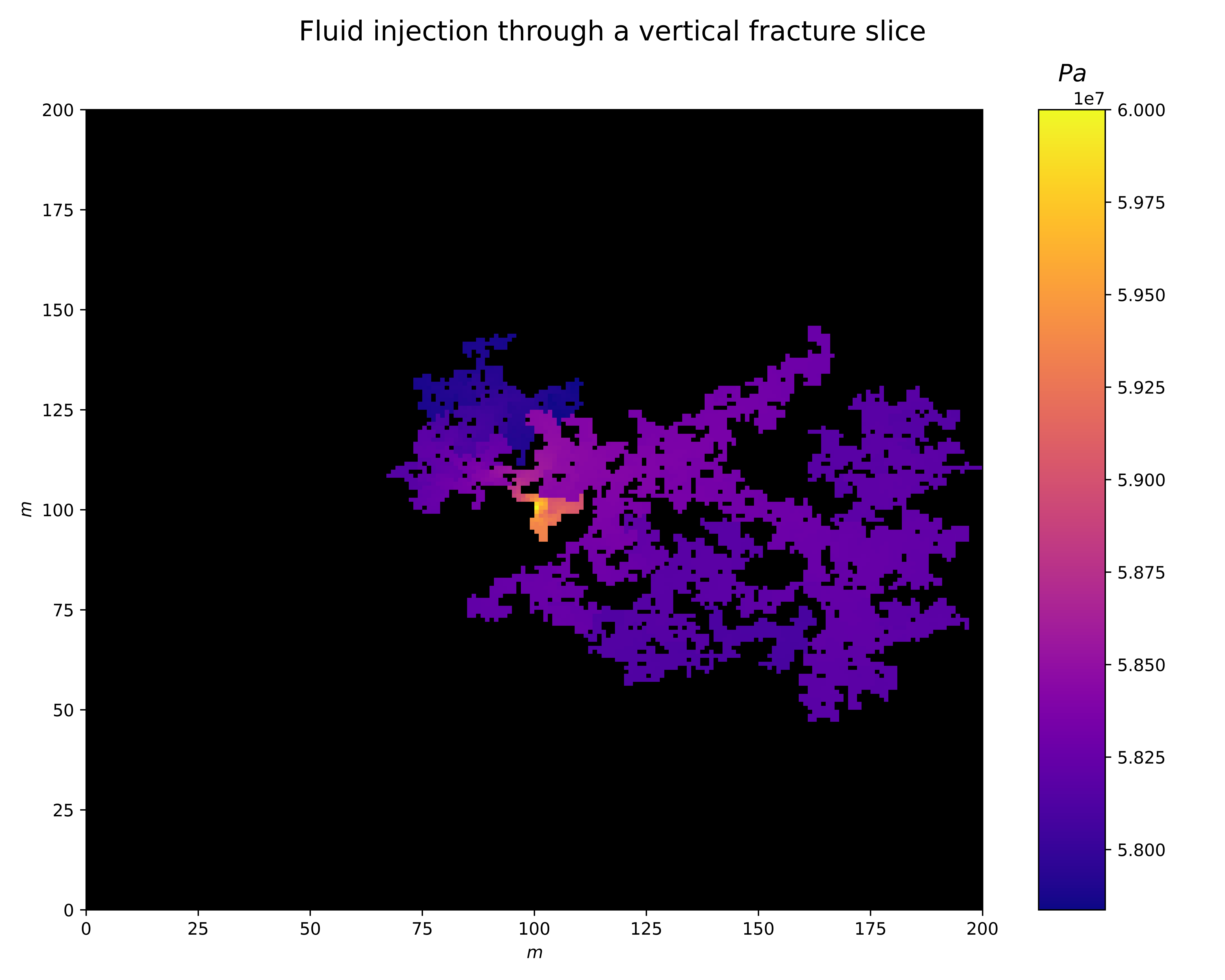}
    \caption{An example injection simulation, with associated spatial pressure drop through the fracture.} 
\end{figure}

This can be demonstrated in the model process for fracking simulation. A fracking fluid injection begins with the occupation of a single cell, with a pressure of $60 MPa$. This is corresponds to the initial pressure, post-breakdown, in a fracking cycle. Fluid begins to flow, mapped to space through invasion percolation. As the fracture expands, the pressure drops further from the borehole. Each pixel of the figure corresponds to a unique inflationary stress source, each with variable moment magnitude tied to pressure.

\begin{equation}
    t_0 = \dfrac{\sum_{i} V_{i}\phi_{i}}{Q_w}
\end{equation}

The total injection time $t_0$ is calculated by summing the fluid held in each occupied cell, by multiplying the volume $V_i$ by the porosity $\phi_i$, and then dividing by the injection well flow rate, $Q_w$.

This injection has immediate effects on local seismic conditions, through the modification of the Green's solutions arrays, but the risk of induced seismicity is seldom from a single injection, but instead repeated, long term, patterns. As a result, time evolution of the fracture is needed. 

The immediate period after shut-in reaches a quasistable state after a short period, but then slowly declines over time. To represent this, after an injection is completed, if a earthquake rupture was not triggered by the injection, the pressures on all nodes in the cluster are set to the mean pressure value of the burst.

Then, the system is iterated in time to the next fluid injection, and each existing cluster of inflationary nodes has its pressure lowered according to the following relations. The pressure decline after shut-in of a fracture in a contained quasielastic space can be described with the following equation\cite{10.2118/8341-MS}.

\begin{equation}
    \Delta P(\delta_0,\delta) = \dfrac{CH_{p}\overline{E} \sqrt{t_{0}}}{{H^2\beta_s}} G(\delta,\delta_0)
\end{equation}

\begin{equation}
    \delta = \frac{\Delta t}{t_o}
\end{equation}

$\delta_{0}$ and $\delta$ are dimensionless variables related to time, a quotient of the elapsed time since shut-in $\Delta t$ and the total injection time $t_0$. $H$ is the height of the fracture itself, measured as the total average vertical range of the occupied cells, referenced from the injection point. $\overline{E}$ is the plane-strain modulus. $\beta_s$ is the ratio of pressure at the wellbore to the average pressure, which is assumed to be $1$ in this case, after local pressure gradients equalize. The $G$ function is described as follows.

\begin{equation}
    \begin{gathered} 
     G(\delta,\delta_0) = \frac{4}{\pi} (g(\delta) - g(\delta_0)) \\
     g(\delta,\delta_0) = \frac{4}{3} ((1+\delta)^\frac{3}{2} - \delta^\frac{3}{2} -1) \\
     \end{gathered}
\end{equation}

$H_{p}$ is the total amount of the fracture that is occupied by proppant, which is the solid material injected with the fluid to keep the fault open during injection. This settles at the bottom after injection ends. The ratio of $H_{p}$ to H varies based on proppant material and injection parameters, but often is in the range .2 to .4\cite{Lin06012025}. Rewriting this with the ratio $\dfrac{H_{p}}{H}$ as $H_r$, results in

\begin{equation}
    \Delta P(\delta_0,\delta) = \dfrac{CH_{r}\overline{E} \sqrt{t_{0}}}{{H\beta_s}} G(\delta,\delta_0)
\end{equation}

The final variable of note, $C$, is the fluid-loss coefficient, which determines the decay rate. Originally an experimentally derived quantity, there have been several theoretical approximations made for numerical simulations.

\begin{equation}
    C = \sqrt{\dfrac{k_rc_t \phi}{\pi \mu_r}} \Delta P_r
\end{equation}

The equation used in this model is listed above\cite{Economides_Nolte_2000}. $c_{t}$ is the compressibility of the fracking fluid. $k_r$, $\phi$, and $\mu_r$ respectively are the permeability and porosity of the unfractured reservoir, and the viscosity of the reservoir fluid. $\Delta P_r$ is the difference in pressure between the fracture and the surrounding reservoir. This depends on the time dependent loss in pressure in the fracture, linking $\Delta P$ and $C$ as differential equations.

The mean pressure value of the injection example was approximately $58 MPa$. The time evolution of this injection cluster, as it approaches the shale oil reservoir pressure is shown below. The sample value for the oil reservoir is $20MPa$.

\begin{figure}[H]
    \centering
    \includegraphics[width=9cm]{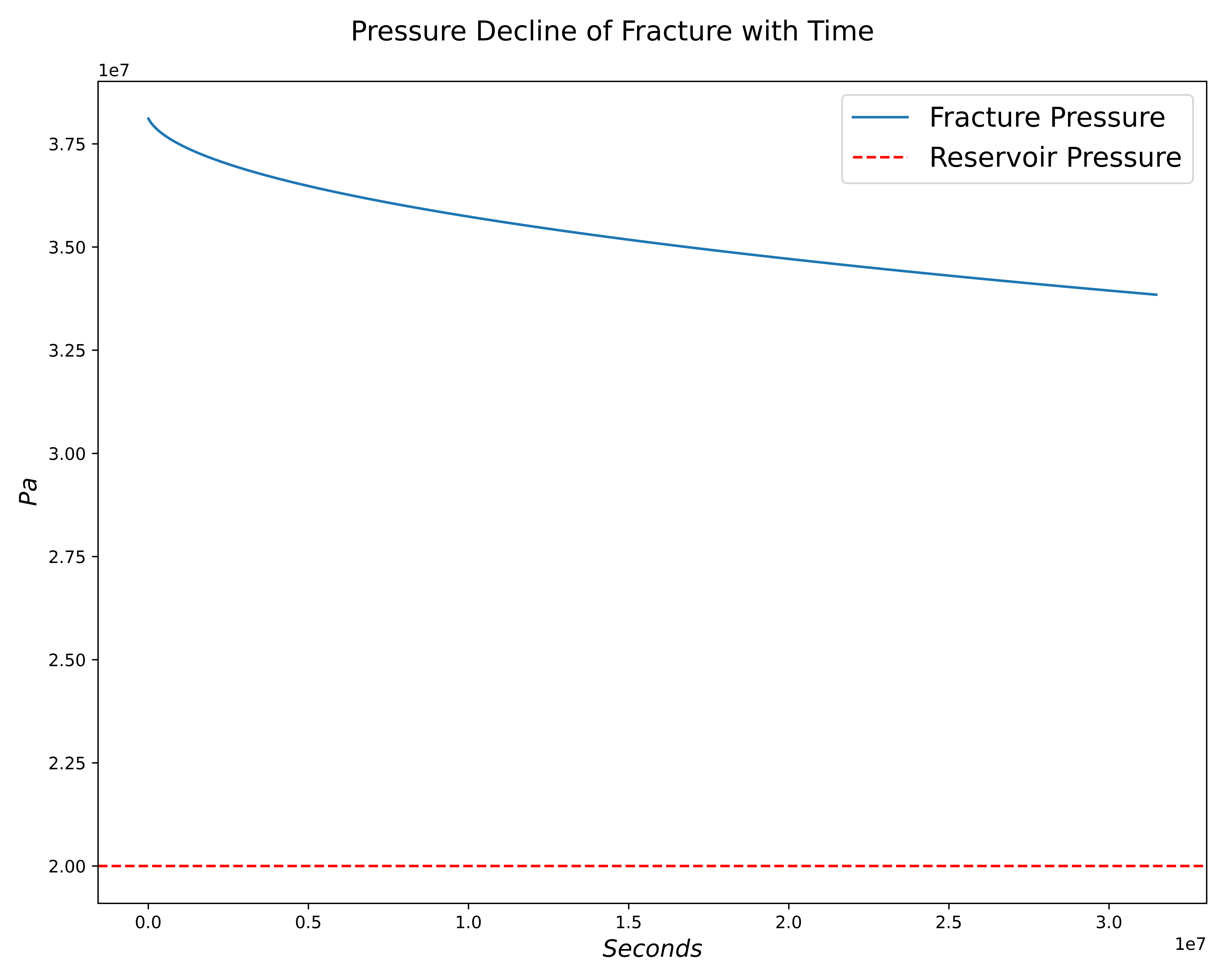}
    \caption{The beginning of pressure equalization over the course of a year, post injection.} 
\end{figure}

This figure shows the pressure over a period of one year from initial injection, and the decline is minimal. While lower than before, the region experiences increased stress over a lengthened period. While one year is small on geologic time scales, it is very long in comparison to the many fracking injections often performed over that period. Repeated fracking has led to larger and larger networks of high pressure fluid, straining faults in ways they would never naturally experience in human time scales.

\begin{figure}[H]
    \centering
    \includegraphics[width=12cm]{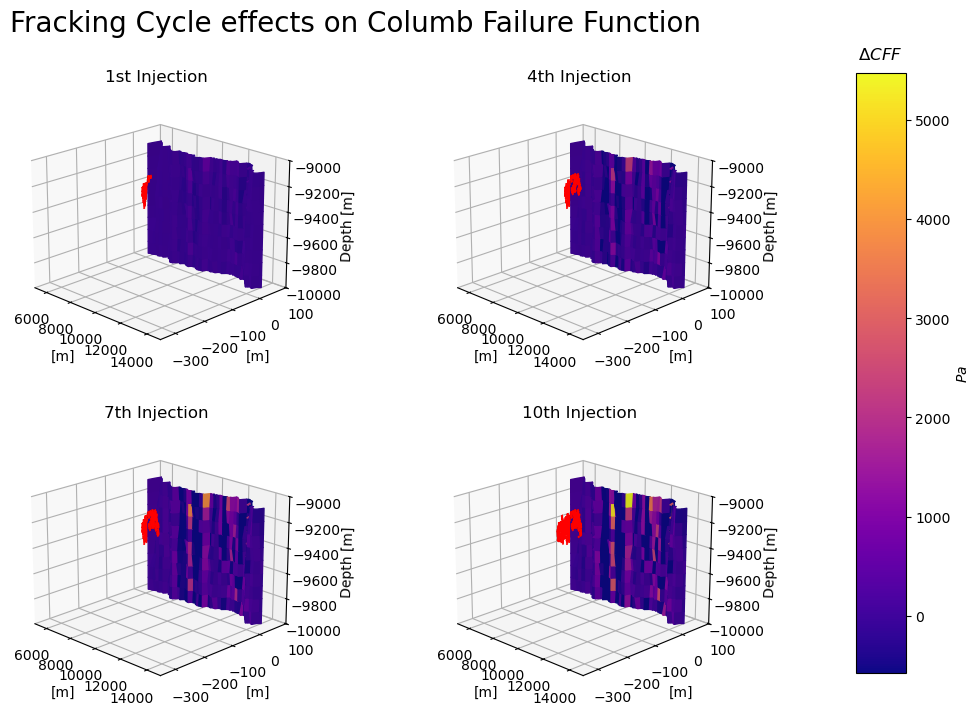}
    \caption{The cumulative effects on $CFF$ from a sample fluid injection. Each voxel square is mapped to an inflationary s} 
\end{figure}
In this figure, a sample of a typical fracking sequence is displayed, an iteration of $10$ fracking cycles. The sequence leads to repeated bursts of injected fluid, the center moving laterally at the given depth. A nearby fault section is rendered, with the net change in the Coulomb Failure Function shown on the surface. While a single injection has only a modest effect on the stresses exerted on a fault system, these sequences happen on a weekly basis in active zones, frequency that is far faster than the time for a pressurized fluid pocket to return normal, background conditions, leading to a progressively less stable fault system and erratic behavior.

This model and its modifications from VirtualQuake seek to provide a way to better understand, and better model these growing disturbances caused by fluid injection. To provide a physics based, instead of statistical method of simulating these interactions before disaster strikes. It has been shown that pausing fluid injections for longer periods allows for local conditions to "cool" and stabilize, a behavior corroborated in the model, and further exploring this simulation space could give better bounds on this process.

As part of this, more effort needs to be done to parameterize the system in regards to real, measured data, to better show the linkage between this proposed model and the real phenomenon affecting communities. Further work will be done to generate training corpora for machine learning models, to show the displacements caused by these injections, and attempt to make predictive tools to better characterize risk.

\printbibliography

\end{document}